\documentclass[12pt,a4paper]{article}
\usepackage[a4paper, total={6in,9in}]{geometry}

\usepackage{amsmath}
\usepackage{amsfonts}
\usepackage{amssymb}
\usepackage{graphicx}
\usepackage{color}
\usepackage{epstopdf}
\usepackage{xcolor,colortbl}
\usepackage{caption}
\usepackage{tikz}
\usepackage{subfigure}

\usepackage{ulem}

\begin{document}

\begin{center}
\begin{Large}
\textbf{
Amplitude Nanofriction Spectroscopy}
\end{Large}
\end{center}

\vspace{0.5cm}

\begin{center}
Antoine Lainé$^{1*}$, Andrea Vanossi$^{2,3}$,  Antoine Niguès$^1$, Erio Tosatti$^{2,3,4**}$ and Alessandro Siria$^1$\\

\vspace{0.5cm}
\textit{$^1$ Laboratoire de Physique de l'\'Ecole Normale Sup\'erieure, ENS, Universit\'e PSL, CNRS, Sorbonne Universit\'e, Universit\'e Paris-Diderot, Sorbonne Paris Cit\'e, UMR CNRS 8550, 24 Rue Lhomond 75005 Paris, France}\\
\textit{$^2$ International School for Advanced Studies (SISSA), Via Bonomea 265, 34136 Trieste, Italy}\\
\textit{$^3$ CNR-IOM Democritos National Simulation Center, Via Bonomea 265, 34136, Trieste, Italy}\\
\textit{$^4$ The Abdus Salam International Centre for Theoretical Physics (ICTP),
Strada Costiera 11, 34151 Trieste, Italy}\\
$^*$ antoine.laine@phys.ens.fr \\
$^{**}$ tosatti@sissa.it
\end{center}

\clearpage
\begin{abstract}
\bf{
Atomic scale friction, an indispensable element of nanotechnology, requires a direct access to, under actual growing shear stress, its successive live phases: from static pinning, to depinning and transient evolution, eventually ushering in steady state kinetic friction. Standard tip-based atomic force microscopy generally addresses the steady state, but the prior intermediate steps are much less explored.  Here we present an experimental and simulation approach, where an oscillatory shear force of increasing amplitude leads to a one-shot investigation of  all these successive aspects. Demonstration with controlled gold nanocontacts sliding on graphite uncovers phenomena that bridge  the gap between initial depinning and large speed sliding, of potential relevance for atomic scale time and magnitude dependent rheology.   
}
\end{abstract}

\clearpage

The advent of nanotechnology demands an understanding of different stages of the time-dependent shearing process of nanoscale contacts, from initial static friction  at low shear stress, to depinning and subsequent evolution towards steady state sliding at large stress.  Among several tools \cite{Persson,  Krim1997}, the Atomic Force Microscope (AFM) \cite{binnig1986} and its variants \cite{Mate1987,Gnecco2007,Dienwiebel2004}  have been a prolific source of data and understanding -- yet mostly concentrated on steady state kinetic friction \cite{Riedo2003,Li2011,Vanossi2013,Park2014}.  Live situations, such as that of nanocontacts that initially stick but subsequently yield and slide with variable dislodgings and unknown Joule dissipation under growing  shear stress, fall outside  that well explored steady state sliding regime.
The present approach is designed to fill that gap, exposing the sequence of  (wear-free) frictional  events incurred by a nanocontact which is shaken with a time-dependent shear protocol. We submit a dry atomic scale contact to an oscillatory time-dependent shear stress of increasing  magnitude and reveal,  by extracting the full complex mechanical impedance, not just the usual frictional dissipation response but also the effective lateral stiffness and its rheological evolution with increasing shear. To that end, we employ a sequence of conductance-controlled gold-graphite nanocontacts, which expose,  in parallel with model simulations,  phenomena typical of each regime, from size-dependent depinning stress — including thermolubric behaviour at small size — to quantized multistep laps,  to the eventual frictional saturation for large amplitude and velocity, eventually dominated by static friction at turning points.     

In a vacuum chamber, a gold tip attached to an in-plane oscillating prong of a Quartz Tuning Fork (QTF) used as a force sensor  is brought in contact with a Highly Oriented Polycrystalline Graphite (HOPG) substrate.  The nanometric longitudinal size of the contact is much smaller than the large electron conduction mean free path ($\sim$ 40 nm) \cite{Haynes2014, Gall2016} across the nanocontact, generally ballistic as opposed to diffusive. 
In that limit, Sharvin’s conductance ~\cite{Wexler1966, Nikolic1999, Erts2000} simplifies to Landauer's form $ G = gG_0 \sim N<T>G_0 $ which   
yields indirect information on the effective contact area  $\phi \sim Nd^2$ where $N$ is roughly speaking the number of atomic chains that constitute the conducting channels,   
$d$ is the transverse distance between them (of order of the lattice spacing), $G_0 = (2e^2) / h$ is the conductance quantum, $e$ the electron charge and $h$ Planck’s constant. In this indirect connection between contact conductance and transverse area,  the average transmission $<T>$  per channel is the main uncontrolled parameter.  From a pure number close to 1 in a metal-metal contact, $<T>$  may vary between 1 and as little as $10^{-2}$, as we shall see below, in a metal-graphite contact \cite{Kim2014}, owing to graphite's very low density of electronic states at the Fermi level, and varying depending upon local contact conditions.  \\

The QTF provides dynamic force measurements within the standard Frequency Modulation AFM framework \cite{Giessibl2019}, with high force resolution. The large mechanical stiffness and subnanometric oscillation amplitude ensure good stability. An imposed sinusoidal force $F^* = F ~ e^{i\omega_0 t}$, at the frequency $f_0=\omega_0/(2\pi) \approx 32$ kHz, leads to a shear oscillation amplitude $a^* = a_{TF} ~ e^{i \left(\omega_0 t + \delta \right)}$ of the prong. The velocity directly reads $v=i\omega_0 a^*$, thus  it is proportional to the oscillation amplitude with a $\pi/2$ dephasing. This force sensor yields the full mechanical impedance of the sliding interface $Z^* = F^* / a^* = Z' + iZ''$ (N/m) with $Z'$ the elastic and $Z''=F_D/a_{TF}$ the dissipative impedance, $F_D$ being the dissipative force (see Supplementary Information section 'Dynamic force measurements'). 
For a broad range of well defined size contacts, from the thinnest, probably monatomic ones with conductance $g \sim 10^{-2}$  to the broadest ones with hundreds/thousands times larger $g$,  our protocol is to extract $Z^*$ under the gradual increase of the oscillation amplitude $a$, an increase actuated while keeping the conductance fixed.
It should be noted  that the large out-of-plane stiffness of our apparatus $k_N \approx 20$ kN/m ensures that any adhesion-driven mechanical instability is \textit{de facto} prohibited. It offers the opportunity to access such atomically small contact area \cite{Sun2020} as compared with cantilever-based AFM experiments, which nicely probed larger contact areas with the advantage of simultaneously measuring the load \cite{enachescu1998}, but are limited by adhesion instabilities. As a consequence, the large out-of-plane stiffness of our experimental design \textit{in fine} enables to obtain the static and dissipative friction forces, at the expense of avoiding any load measurement and subsequent friction coefficient estimation so far.  

\begin{figure}[h!]
\centering
\includegraphics[width=0.8\columnwidth]{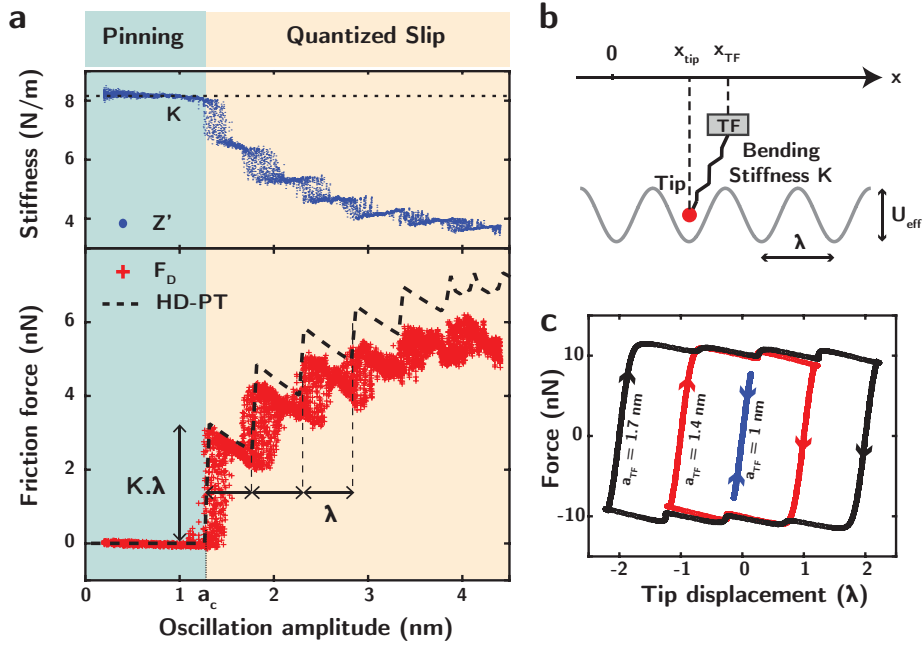}
\caption{\textbf{Amplitude spectroscopy} 
\textbf{a} Measured effective lateral stiffness $Z' = 2k_{TF} (\delta f/f_0)$ and frictional force $F_D = Z" a_{TF}$  of  a gold tip-graphite nanocontact held at a fixed average conductance $g\sim 1$ as a function of increasing oscillation amplitude $a$.   Dashed line: simulated frictional force $F=\oint K(x_{TF}-x_{tip}) dx_{tip}$ for a HD-PT model with parameters  $\lambda = 0.4$ nm, $T=0$ K, $K = 8$ N/m, $U_{\text{eff}} = 14$ eV, $m = 10^{-13}$ kg, $\mu = 2\sqrt{K/m} \approx 0.02$ ns$^{-1}$).
Note the transition from the pinning regime (green, constant $Z'=K$ and $F_D=0$) to successively quantized slips (orange, $Z'$ and $F_D$ respectively decreasing and increasing by discrete jumps, with good agreement between data and simulation.) 
\textbf{b} Schematic PT model. \textbf{c} Simulated quantized slips give rise to friction loop openings for increasing oscillation amplitude,. $F(t) = K \cdot \left( x_{TF}(t) -x_{tip}(t) \right)$ plotted as a function of normalized tip displacement $x_{tip}/\lambda$. Evolution from pure elastic regime (blue), to sliding over 3 potential minima (red) at  $(-\lambda,0,\lambda)$ , then over 5 potential minima (black) at $(-2\lambda,-\lambda,0,\lambda,2\lambda)$.
}
\label{fig:Qslip} 
\end{figure}

Fig. \ref{fig:Qslip} displays a typical result of our amplitude frictional protocol for a medium size contact, $g \sim 1$. Mechanical impedance is purely elastic up to the depinning {\it static friction}
oscillation amplitude $a_c$, and dissipative above that, with  jump amplitudes close to $\lambda= 2 \times a \sqrt{3}= 0. 49$ nm, where $a =  0.246$ nm is graphite’s lattice spacing. That suggests that jumps can be attributed to multi-valley $[-(n\lambda), +(n\lambda)] \longleftrightarrow [-(n+1)\lambda, +(n+1)\lambda]$ slip events, with $n=1,2,$ … as the oscillating contact samples graphite’s corrugation minima after the initial pinning at $n=0$. 

To get insight into the contact frictional evolution as the shear amplitude grows, we conduct a parallel simulation of a 1D  harmonically driven Prandtl-Tomlinson (HD-PT) model \cite{Dudko2002, Dong2011} (Fig. \ref{fig:Qslip} \textbf{b}) whose  equation of motion is numerically solved for the tip apex dynamics $x_{tip}(t)$, while imposing a harmonic external drive $x_{TF}(t)= a_{TF} \cdot \text{cos}(\omega_0 t)$, initially assuming $T=0$ K (see Supplementary Information section 'Harmonically driven PT model - Description and computation'). 
The simulated instantaneous force acting on the tip as a function of the normalized tip displacement $x_{tip}(t)/\lambda$ (Fig. \ref{fig:Qslip} \textbf{c}) describes the transition from a pinned elastic regime to the unpinned dissipative regime exhibiting discrete friction jumps while increasing the model oscillation amplitude. In the pinned regime, the displacement is accommodated by a purely elastic deformation (blue, linear and fully reversible force) such that the tip apex remains in the initial potential minimum $x_{tip}/ \lambda \approx 0$ and the dissipation is basically null. Increasing the driving oscillation amplitude (red, $a_{TF} = 1.4$ nm) above a static friction threshold provokes depinning followed by frictional sweeps with  back and forth swinging via slip events between the turning points $(-\lambda, 0, \lambda)$. A hysteresis loop opens up in the force response whose area corresponds to energy dissipation per cycle. 
By  increasing the oscillation amplitude (black, $a_{TF} = 1.7$ nm), the friction loop increases over an increasing number of substrate's potential minima $(-2\lambda,-\lambda, 0, \lambda,2\lambda)$, etc. 
The good fit to the experimental impedance data (black dashed line in Fig. \ref{fig:Qslip} \textbf{a})  permits the quantitative determination of the effective PT model parameters: stiffness $K$, substrate's spatial periodicity $\lambda$ and corrugation $U_{\text{eff}}$. From the expeimental depinning amplitude $a_c$, the dimensionless parameter $\eta_{expt}=2\pi a_c/\lambda$ can  also be extracted (see Supplementary Information section 'Harmonically driven PT model - Static parameters determination from depinning'), a quantity that in the PT model must equal $\eta=(2\pi^2U_{\text{eff}})/(K \lambda^2)$ separating the friction regimes between  smooth sliding  ($\eta<1$) and stick-slip ($\eta>1$) \cite{Socoliuc2004}. The medium size contact with  $g \sim 1$ of  Fig. \ref{fig:Qslip} has  $\eta \sim 50$, far in the stick-slip motion regime.    

Contacts of decreasing conductance have a smaller and smaller static friction threshold  $a_c$. While as we shall explain below all contact sizes  should still correspond to $\eta>1$,  a finite threshold  $a_c$ rapidly becomes undetectable in the smallest contacts. That probably involves an important role of thermal fluctuations, a point to which we shall return  further below. The blurring of the experimental force response at each jump is also likely to be due to thermally induced barrier hopping events at the turning points of the oscillation cycle. Indeed, at the turning point, the tip velocity goes to zero and the effective energy barrier may become small with respect to thermal energy (see Supplementary Information 'Turning point thermolubricity'). In the following we analyze the different regimes and phenomena that have become accessible.

{\it Effective elastic stiffness drops.} Simultaneously to the dissipative jumps, the quantized slip cycles open up, with a sequence of mechanical softening events that can be described as drops of the effective lateral average stiffness $\langle Z' \rangle = f_0 \int_0^{1/f_0} Z'(t) ~ dt$ under increasing oscillation amplitude.
At small amplitudes below depinning, $\langle Z' \rangle = K$,  $\langle Z"\rangle \sim 0$. Conversely, above the depinning amplitude  $a_c$, as frictional slips kick in (orange regions in Fig. \ref{fig:Qslip} \textbf{a}), the averaged stiffness concomitantly drops. There is in fact a perfect correlation between the dissipative force jump and the stiffness drop within each dissipative steps (see Supplementary Information section 'Stiffness drop and dissipation increase correlation'), underlining the common origin of the two observed features. The instantaneous stiffness vanishes during slip events. Thus, as the slip number within an oscillation cycle becomes larger, the averaged stiffness  $\langle Z' \rangle$  decreases with increasing amplitude.

\begin{figure}[h!]
\centering
\includegraphics[width=0.5\columnwidth]{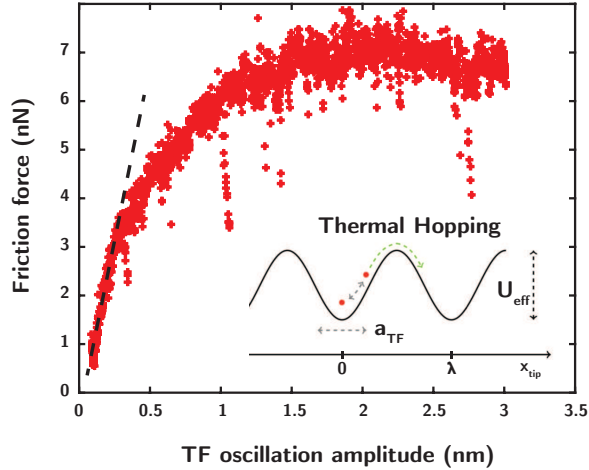}
\caption{\textbf{Thin contact thermally assisted lubricity regime} Amplitude spectroscopy of a thin contact interface for $g \sim 0.01$. Disappearance of the observed static friction and linear decrease of the dissipation force with velocity going to zero. Inset shows a schematic of a thermal hopping event.}
\label{fig:sCor} 
\end{figure}

{\it Atomically small contacts: thermolubricity }. Amplitude spectroscopy of the thinnest contacts reveals -- see for example for $g \sim 0.01$ in Fig. \ref{fig:sCor} -- the disappearance of static friction, with apparently pinning-free sliding down to nominally zero amplitude.  That is unexpected in  $T=0$ mechanics where, theoretically, the static friction should never vanish, 
since the starting point at zero amplitude is a local energy minimum, protected by a  barrier. The vanishing static friction shown in Fig. \ref{fig:sCor}  must therefore be ascribed to thermal fluctuations across these barriers (see inset in Fig. \ref{fig:sCor}). Very well known in the PT model \cite{Muser2011, Pellegrini2019}  and also documented in trapped cold ion sliding ~\cite{Gangloff2015}  this  thermally assisted crossover from stick-slip to viscous sliding, sometimes called "thermolubricity" \cite{Krylov2014},  is not commonly reported in nanofriction experiments. An additional demonstration of thermolubricity in our smaller size  contacts ($g < 0.5$) besides the demise of static friction is offered in  Fig. \ref{fig:sCor} by the linear drop of frictional dissipation force $F_D$, indicating viscous friction,  in the limit of vanishing  amplitude \cite{Pellegrini2019}.

\begin{figure}[h!]
\centering
\includegraphics[width=0.5\columnwidth]{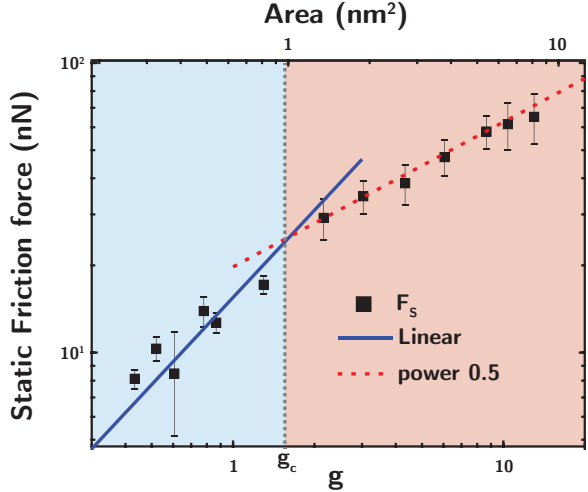}
\caption{\textbf{Atomic-scale static friction} Static friction force as a function of conductance (in turn proportional to contact area, upper axis). By increasing the contact area, note the transition from linear (blue) to sublinear power law (red) regime. The dashed line in the sublinear regime indicates a power of about 0.5.  
Error bars show the standard error.
}
\label{fig:Flaws} 
\end{figure}

{\it Large contacts: bulk to edge friction } For larger size contacts, the static friction threshold amplitude $a_c$ grows with growing conductance $g$, also connected to the static friction force $F_S = K~a_c$. Thus, varying the conductance $g=N<T>$, we extract the evolution of the static friction with the contact area $\phi = N \phi_0$, where $\phi_0 = \pi d^2$ denotes the single atom effective area.  The evolution of the contact stiffness with conductance (see Supplementary Information section 'Stiffness-conductance evolution') evidences the linear dependence between conductance and contact area within the range of interest. Assuming, arbitrarily but reasonably, $<T> \approx 0.1$ following \cite{Kim2014} we build Fig. \ref{fig:Flaws}. The static friction force covers two different regimes as the contact area increases. For a smaller contacts  $g<g_c \approx 1.5$ (blue region, Fig \ref{fig:Flaws}), corresponding to a area $\phi_c \approx 1$ nm$^2$ made of $N_c\approx 5$ atoms (taking $d \approx 0.25$ nm as the typical lattice spacing), the static friction force increases linearly with the contact area $F_S \approx \sigma_S \phi$ with a shear stress $\sigma_S \approx 5$ nN/atom.
For larger  areas instead, $\phi > \phi_c$, the static friction growth becomes sublinear (red region, Fig. \ref{fig:Flaws})  $F_S \sim (\phi/\phi_0)^{\alpha}$, with $\alpha \approx 0.5$. The transition between these two regimes is reminiscent of the friction crossover from proportionality to area to perimeter, the latter taking place for larger contacts, either incommensurate where edge effects dominate \cite{Varini2013}, or amorphous where random asperities dominate~\cite{Muser2001}. For $g<g_c$, the contact is small enough that all atoms contribute to the static friction while for $g>g_c$, only the fraction of atoms near edges or asperities participate in the pinning process.

\begin{figure}[h!]
\centering
\includegraphics[width=\columnwidth]{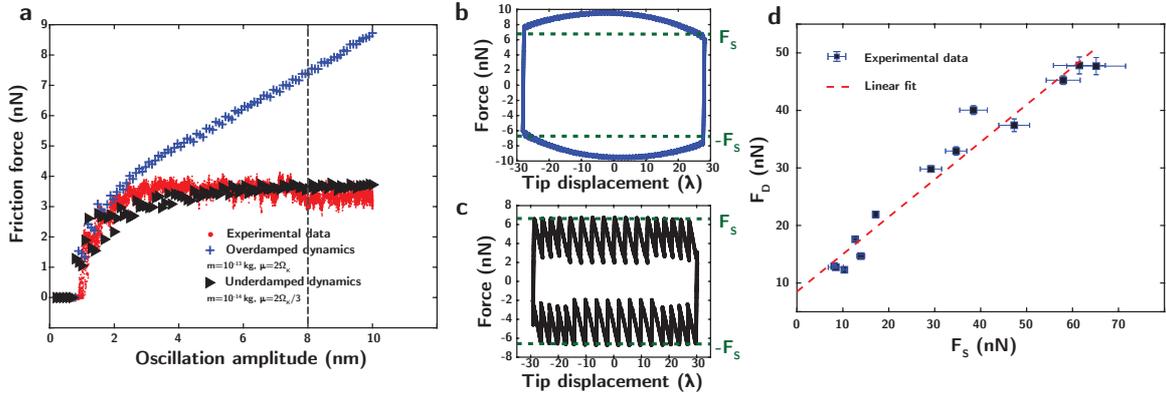}
\caption{\textbf{Multiple slip at large oscillation amplitude} \textbf{a} Solid-like large amplitude dissipation. Evolution of the dissipative friction force as a function of TF oscillation amplitude. Experimental results (red dots) and numerical results in the overdamped (blue crosses) and underdamped (black triangles) regimes. \textbf{b} and \textbf{c} Friciton loops in the overdamped and underdamped regime respectively. \textbf{d} Linear dependence of the dissipative friction force as a function of static friction force.
Error bars show the standard error.
}
\label{fig:LA} 
\end{figure}

{\it Large amplitude: macroscopic friction} At large oscillation amplitude, the dissipation force tends towards a macroscopic, solid-like frictional regime characterized by a kinetic friction  $F_D^{\infty}$ essentially independent (Coulomb's law) of velocity,  here proportional on average to the oscillation amplitude (Fig. \ref{fig:LA} \textbf{a}, red points). 
Most of the frictional loop, corresponding to the fast $(-n\lambda, n\lambda)$ swing, is  effectively viscous and, owing to small damping, poorly dissipative. On the other hand, the sliding contact comes to a stop, and becomes pinned, twice per cycle at the turning points: and the subsequent two depinnings are the major source of dissipation. Simulation results highlight this feature in Fig. \ref{fig:LA} \textbf{b} and \textbf{c} for both overdamped and underdamped dynamics. For the overdamped case (Fig. \ref{fig:LA} \textbf{b}), the instantaneous force overcomes the static friction value within the tip oscillation cycle, thus providing extra dissipation which leads to the linear increase of the dissipation with applied oscillation amplitude (Fig. \ref{fig:LA} \textbf{a}, blue crosses). The high damping leads to single slip events compelling the tip to reach a potential minima closer to the global minima. However, in the underdamped case, the static friction represents an upper value for the instantaneous force (Fig. \ref{fig:LA} \textbf{c}) and leads to dissipation in accordance with the experimental results. We conclude that the large-amplitude macroscopic friction is intimately related to the underdamped nature of the tip dynamics. This is consistent with the occurence of multiple slip events as observed in Fig. \ref{fig:LA} \textbf{c}.

{\it Large amplitude: kinetic vs. static  friction} If indeed originated by the depinning events at the turning points, the measured large-amplitude kinetic  friction  $F_D^{\infty}$ should therefore be controlled by the static friction $F_S$. By varying the contact area, we vary the static friction (see Fig. \ref{fig:Flaws}) and simultaneously observe the area dependence of the large amplitude kinetic friction . Fig. \ref{fig:LA} \textbf{d} confirms a linear connection between kinetic and static friction in the large amplitude regime. A parallel analytical estimate of the expected kinetic dissipation as a function of the static friction force in the underdamped case yields $F_D = \xi F_S$, with $\xi_{1S} = 1$ for single slip events and $\xi_{MS} = 0.5$ for the largest possible multiple slip (see Supplementary Information section 'Large amplitude dissipation from static friction'). The experimental result yields $\xi \approx 0.65$, thus pointing towards the intermediate case of an incomplete multiple slip dynamics. The large $\eta$ values ensures the accessibility of multiple slip \cite{Medyanik2006}. Similarly the large velocity accessed with our experimental schemes favors multiple slip events \cite{Gangloff2015}. The limited multiplicity of the slip is therefore very likely to arise from the actual damping value.\\

In summary, by accessing, in an increasing amplitude mode, the complex mechanical impedance during oscillatory shear of conductance-controlled nanocontacts in an otherwise standard tuning fork AFM, we show how a multitude of relevant time dependent frictional phenomena that emerge can be addressed, as it were, in one shot. The static friction, which at small areas is found to vanish, demonstrating thermolubricity, first then grows linearly with size, eventually crossing over at large sizes to the edge-dominated sublinearity expected from incommensurate superlubricity. The series of discrete depinning events observed for increasing amplitude correspond to quantized expansions of the multiple-slip swings between further and further equivalent minima. The kinetic friction, initially growing fast with amplitude (and thus with velocity), eventually settles to a weak to negligible growth, showing that its origin is mainly in static friction - here at the oscillation turning points - exactly as in macroscopic solid sliding. While much of this behaviour can, as befits a demonstration such as the present one, be rationalized by a parallel Prandtl-Tomlison model simulation, the experimental approach, which extends previous ones \cite{Bylinskii2015,Wada2018}, has the ability to reach unprecedented velocities and other parameters, offering a considerably wider potential for probing friction at the atomic scale and exploring novel atomic-scale dissipation concepts, of growing importance in nanotechnology. 

\bibliographystyle{ieeetr}
\bibliography{Arxiv_ANS}

\end{document}